\documentclass[10pt,preprint]{aastex}
\shorttitle{The Structure of the Homunculus}
\shortauthors{Smith \& Ferland}
\begin{document}

\title{THE STRUCTURE OF THE HOMUNCULUS. II.  MODELING THE PHYSICAL
  CONDITIONS IN ETA CAR'S MOLECULAR SHELL\altaffilmark{1}}

\author{Nathan Smith\altaffilmark{2,3}}
\affil{Center for Astrophysics and Space Astronomy, University of
Colorado, 389 UCB, Boulder, CO 80309}

\author{Gary J.\ Ferland} 
\affil{Department of Physics and Astronomy, University of Kentucky,
Lexington, KY 40506; gary@pa.uky.edu}

\altaffiltext{1}{Based in part on observations obtained at the Gemini
Observatory, which is operated by AURA, under a cooperative agreement
with the NSF on behalf of the Gemini partnership: the National Science
Foundation (US), the Particle Physics and Astronomy Research Council
(UK), the National Research Council (Canada), CONICYT (Chile), the
Australian Research Council (Australia), CNPq (Brazil), and CONICET
(Argentina).}

\altaffiltext{2}{Hubble Fellow}

\altaffiltext{3}{Current address: Astronomy Department, 601 Campbell
  Hall, University of California, Berkeley CA 94720;
  nathans@astro.berkeley.edu}

\begin{abstract}

We present models that reproduce the observed double-shell structure
of the Homunculus Nebula around $\eta$~Carinae, including the
stratification of infrared H$_2$ and [Fe~{\sc ii}] emission seen in
data obtained with the Phoenix spectrograph on Gemini South, as well
as the corresponding stratified grain temperature seen in
thermal-infrared data.  Tuning the model to match the observed shell
thickness allows us to determine the threshold density which permits
survival of H$_2$.  An average density of
$n_H\simeq$(0.5--1)$\times$10$^7$ cm$^{-3}$ in the outer zone is
required to allow H$_2$ to exist at all latitudes in the nebula, and
for Fe$^+$ to recombine to Fe$^0$.  This gives independent
confirmation of the very large mass of the Homunculus, indicating a
total of roughly 15--35 M$_{\odot}$ (although we note reasons why the
lower end of this range is favored).  At the interface between the
atomic and molecular zones, we predict a sharp drop in the dust
temperature, in agreement with the bimodal dust color temperatures
observed in the two zones.  In the outer molecular shell, the dust
temperature drops to nearly the blackbody temperature, and becomes
independent of grain size because of self-shielding at shorter UV
wavelengths and increased heating at longer wavelengths.  This relaxes
constraints on large grain sizes suggested by near-blackbody color
temperatures.  Finally, from the strength of infrared [Fe~{\sc ii}]
emission in the inner shell we find that the gas-phase Fe abundance is
roughly solar.  This is astonishing in such a dusty object, where one
normally expects gaseous iron to be depleted by two orders of
magnitude.

\end{abstract}

\keywords{circumstellar matter --- ISM: dust extinction --- ISM:
  individual (Homunculus nebula) --- stars: individual ($\eta$
  Carinae) --- stars: winds, outflows}

\section{INTRODUCTION}

The bipolar Homunculus Nebula surrounding $\eta$ Carinae offers an
unusually well-constrained laboratory to study a wide array of
physical processes in the interstellar medium.  The distance, size,
shape, orientation, and structure of the nebula are all known (Smith
2006; Davidson et al.\ 2001), as are its expansion velocity and age
(Morse et al.\ 2001; Currie \& Dowling 1999; Smith \& Gehrz 1998).
Ionized ejecta outside the Homunculus indicate nitrogen-rich gas
(Davidson et al.\ 1986; Dufour et al.\ 1997; Smith \& Morse 2004), and
the recent detection of ammonia in the Homunculus suggests that it too
is N-rich (Smith et al.\ 2006).  The Homunculus itself is mostly
neutral, since the star's dense wind extinguishes nearly all of the
Lyman continuum produced by the system.  This means that familiar
temperature and density diagnostics of low-density ionized gas in
H~{\sc ii} regions (e.g., Osterbrock \& Ferland 2006) cannot be used
here.  At UV/visual wavelengths the Homunculus is seen mainly in
reflection.  The dust that absorbs and scatters this light acts as a
calorimeter by re-emitting the absorbed luminosity in the infrared
(IR), providing us with a practical and accurate measure of the
bolometric luminosity of the central source (e.g., Smith et al.\
2003b; Cox et al.\ 1995; Hackwell et al.\ 1986; Davidson 1971; Pagel
1969; Westphal \& Neugebauer 1969).  Despite these constraints, $\eta$
Car's nebula is such a bright and complex object that it leaves no
shortage of observational puzzles.  These can sometimes make it
difficult to see the trees for the forest.

Relevant physical conditions in the Homunculus nebula can be
constrained if we understand the origin of the distinct double-shell
structure in the walls of the nebula.  These layers manifest
themselves in three independent ways that have been clearly observed.
First, high resolution thermal-IR images reveal a stratification in
the dust color temperature, with a thin outer shell at $\sim$140 K,
and a warmer, thicker inner zone with a color temperature of $\sim$200
K (Smith et al.\ 2003b).  Second, near-IR spectra show an outer skin
of H$_2$, and an inner shell in tracers of dense, low-ionization
atomic gas like collisionally-excited lines of [Fe~{\sc ii}] (Smith
2006, 2002a).  These two layers are segregated from one another.
Third, the two distinct shells can be seen at different radial
velocities in UV absorption profiles along our line-of-sight to the
star (Gull et al.\ 2006, 2005; Nielsen et al.\ 2005). The outer
molecular layer is also detected in UV absorption lines of CH and OH
(Verner et al.\ 2005), and the inner layer is also seen in
visual-wavelength emission lines like [Ni~{\sc ii}] and [Fe~{\sc ii}],
plus faint emission of H$\alpha$ and [N~{\sc ii}] (Davidson et al.\
2001; Hillier \& Allen 1992; Allen \& Hillier 1993; Smith et al.\
2003a).  A detailed account of this structure was given recently in
Paper I (Smith 2006).

It is this pronounced double-shell structure that we aim to reproduce
in the model presented here.  We are particularly interested in the
physical conditions that allow H$_2$ to form and survive in such a
thin layer.  So far, $\eta$ Carinae is the only luminous blue variable
(LBV) known to be surrounded by molecular gas in its own ejecta, while
nearly all LBV nebulae exhibit bright near-IR emission of [Fe~{\sc
ii}] (Smith 2002b; Smith \& Hartigan 2006).  The H$_2$ is probably a
consequence of the extreme youth and high density of the nebula.  New
observations presented in Paper I, obtained with the Phoenix
spectrograph on the Gemini South telescope, have shown the H$_2$ layer
to be surprisingly thin (only a few per cent of the radius) and
uniformly so, despite the fact that it traces a wide range of
different radii at various latitudes in the bipolar nebula.  This
hints that the near-IR H$_2$ emission arises at the inside surface of
an optically-thick shell, which absorbs the remainder of whatever FUV
radiation penetrates the inner [Fe~{\sc ii}]-emitting zone.  By
varying the density in this outer shell, we can test what physical
density permits H$_2$ to survive.  This density constraint, in turn,
provides an independent measure of the mass of the Homunculus because
the geometric thickness of the shell is known.  The total mass is
thought to be at least 12 M$_{\odot}$ derived from IR dust
observations (Smith et al.\ 2003b).\footnote{Note that the gas:dust
mass ratio of 100 assumed in deriving this mass might be too low, due
to the depletion of C and O in the ejecta.  Also, $\eta$ Car's grains
may have unusual composition (Chesneau et al.\ 2005).}  Here we are
fine-tuning an earlier uniform-density shell model for the physical
conditions in the Homunculus (Ferland et al.\ 2005) in order to
account for the observed thin structure of the H$_2$ layer.

\section{ADOPTED PARAMETERS IN CLOUDY}

We simulate the conditions within the nebula using the development
version of {\sc cloudy}, last described by Ferland et al. (1998).  The
incident stellar continuum is represented by an interpolated 20,000~K
CoStar atmosphere with a total luminosity of 5$\times$10$^{6}
L_{\odot}$. To account for the hard X-rays from the colliding wind
binary system, we add a high-energy component corresponding to a
3$\times$10$^{6}$ K blackbody with a luminosity of 30 $L_{\odot}$
(e.g., Corcoran et al. 2001). The lack of a prominent H~{\sc ii}
region shows that few hydrogen-ionizing photons strike the inner edge
of the nebula, most likely due to absorption by the stellar
wind. Therefore, we extinguish the net continuum by photoelectric
absorption from a neutral hydrogen layer of 10$^{21}$ cm$^{-2}$ to
account for this. Some high-energy photons are transmitted and they
help drive the chemistry.  We also include the galactic background
cosmic ray ionization rate, although the actual ionization rate may be
higher if radiative nuclei are present. Cosmic rays have effects that
are similar to X-rays, providing ionization that helps drive the
chemistry.  We adopt solar gas phase abundances except for the
following: He/H=0.4, [C/H]=--4.4, [N/H]=--2.82, and [O/H]=--4.12.  The
gas:dust mass ratio was set to the normal ISM value of 100, and we
adopted silicate grains rather than graphitic grains because the 2175
\AA\ absorption feature is absent in observations (e.g., Hillier et
al.\ 2006) and the 10~$\micron$ silicate emission feature is strong
(see, however, Chesneau et al.\ 2005).

We chose the nebular geometry to be a thick shell separated into two
radial zones, where the concentric shells are permitted to have
different densities.  Table 1 gives our adopted geometric and physical
properties of the double-shell structure.  The inner shell (Zone 1)
has lower density and is geometrically thicker, corresponding to the
gas traced by infrared [Fe~{\sc ii}] lines and warm $\sim$200 K dust.
The outer shell (Zone 2) is very thin, and corresponds to the material
traced by H$_2$ emission and cooler $\sim$140 K dust.  The outer edge
of the inner shell is directly in contact with the inner surface of
the outer shell.  We calculated the physical conditions in these two
zones separately in {\sc cloudy}, using the {\it transmitted}
continuum at the outside of Zone 1 as the {\it incident} continuum for
Zone 2.  Although the shells were calculated using spherical symmetry
in {\sc cloudy}, they are meant to approximate the observed properties
of the nebula roughly along our sight line through the polar region of
the bipolar lobes.  The equatorial region of the nebula with much
smaller radii was modeled separately (see below).  The inner and outer
radii for the two zones were taken directly from observations (Paper
I) and are not free parameters.

Our goal is to investigate the densities in these two zones, to answer
the following questions: 1) What is the density in Zone 1 that is
consistent with the observed $n_e$ value of $\la$10$^4$ cm$^{-3}$
(Smith 2002a) and permits significant FUV radiation to fully penetrate
the shell, causing iron to remain as Fe$^+$ throughout? 2) What is the
minimum density in Zone 2 that allows hydrogen to become predominantly
molecular, and iron to recombine to Fe$^0$?  The most important free
parameters here are therefore the value of $n_H$ chosen for each zone,
but other factors like the gas:dust mass ratio and the filling factor
play roles as well.  The main determining factor in whether or not
H$_2$ can survive in Zone 2 is the attenuation of FUV radiation in the
Lyman and Werner bands by the total column density of gas and dust in
both zones.  Based on spatially-resolved infrared [Fe~{\sc ii}]
spectra (Paper I) we adopt a volume filling factor of 0.5 in Zone 1.
From similar H$_2$ spectra and the size scales of structures seen in
{\it HST} images (Morse et al.\ 1998) we adopt a filling factor of 0.9
for the thin outer shell in Zone 2.  A fraction of 0.9 also matches
the fraction of the total stellar luminosity absorbed by dust grains
that is reemitted in the IR, while $\sim$10\% of the starlight escapes
at near-UV and visual wavelengths.

Finally, we note that we have ignored the possible influence of the
Little Homunculus (Ishibashi et al.\ 2003).  This inner nebula has a
much lower mass -- a few percent of the Homunculus (Smith 2005) -- and
is apparently clumpy on large scales, allowing both radiation and wind
to leak through. Thus, its influence should be minimal, absorbing
whatever small amount of Lyman continuum radiation may escape the
dense stellar wind, causing strong variability in the radio continuum
(Duncan et al.\ 1997).

\section{RESULTS ALONG OUR SIGHT-LINE}

Figure 1 shows the results of a {\sc cloudy} calculation matched to
observations of the polar lobes of the Homunculus roughly along our
line-of-sight (Paper I).  The transition between Zones 1 and 2 is
sudden at a depth of a little more than 7$\times$10$^{16}$ cm, showing
a thick inner zone traced by [Fe~{\sc ii}] and a very thin outer shell
seen in H$_2$ emission.  In order to reproduce this sharp transition,
the simulation required a strong density increase of a factor of
15--30 from Zone 1 to 2.

Throughout Zone 1, hydrogen remains fully atomic, while Fe is singly
ionized (representative values about halfway through Zone 1 give an
H$_2$ fraction of 5$\times$10$^{-10}$ and a neutral Fe fraction of
2$\times$10$^{-6}$).  The electron density from infrared [Fe~{\sc ii}]
line ratios in Zone 1 is n$_e \la$10$^4$ cm$^{-3}$ (Smith 2002a).  To
match this constraint, the hydrogen density in Zone 1 was set to
10$^{5.5}$ cm$^{-3}$, since n$_e$/n$_H$ was only about 3\% at this
distance from the star.  Only 1--2\% of the hydrogen was ionized at
the innermost edge of Zone 1, probably through Balmer continuum
ionization from the 2s level.  This may account for the faint
H$\alpha$ emission there (Smith et al.\ 2003a; Davidson et al. 2001),
as well as faint radio continuum (Duncan et al.\ 1997).  In both Zone
1 and 2, heating of the gas was dominated ($\ga$90\%) by grain-gas
photoelectric heating, whereas direct heating by line absorption was
minimal.

To investigate the formation of H$_2$ and the recombination of Fe in
Zone 2 we considered several different values for the hydrogen
density, and discuss a few representative values here: n$_H$=10$^6$,
10$^{6.3}$, 10$^{6.5}$, 10$^{6.7}$, and 10$^7$ cm$^{-3}$.  Table 1
lists the properties in Zone 2 at these various densities, and Figure
2 gives a detailed view of the transition to molecular gas for each of
these adopted densities.

With n$_H$=10$^6$ cm$^{-3}$, we found that H$_2$ did not form at all,
with H remaining fully atomic and Fe remaining singly ionized
throughout Zone 2.  When the density was raised to 2$\times$10$^6$
cm$^{-3}$, H$_2$ did begin to form at the outer edge of Zone 2; the
molecular fraction reached a maximum of about 40\% at the outermost
edge (Fig.\ 2).  However, Fe remained singly ionized throughout Zone 2
at this density.  This is not in agreement with observations, since
Fe~{\sc i} absorption is seen from Zone 2 in UV spectra (Nielsen et
al.\ 2005), so the density must be higher than 2$\times$10$^6$
cm$^{-3}$.  At a density of 3$\times$10$^6$ cm$^{-3}$, hydrogen became
99\% molecular at the outer edge of the shell (Fig.\ 2), but Fe$^+$
still did not recombine to Fe$^{0}$.

Observations suggest that H$_2$ should form and the recombination of
Fe should occur {\it early} in Zone 2, especially at the remote polar
regions where the radiation field is weaker than at other latitudes.
This condition is met only for higher hydrogen densities around
10$^{6.7}$ to 10$^7$ cm$^{-3}$ (Fig.\ 2).  At a depth of only about
10$^{15}$ cm into Zone 2 (10--20\% of the way through the layer), the
H$_2$ fraction begins to rise sharply.  This is because its
destruction rate collapses as UV radiation in the Lyman and Werner
bands becomes optically thick and is extinguished by the much higher
column density of gas and dust encountered here.  Once molecules begin
to form, the transition is rapid.  Hydrogen remains fully molecular
throughout the rest of Zone 2, while Fe$^+$ does finally recombine.
The Fe$^+$/Fe$^0$ transition occurs at a slightly larger depth than
the atomic/molecular hydrogen transition, because Fe$^0$ can still be
ionized by Balmer continuum photons with wavelengths longer than the
Lyman-Werner bands.  The hydrogen density is probably not much above
10$^7$ cm$^{-3}$, because then the H$_2$ shell would be thicker than
observed, and the extinction to the star would be unreasonably high.

The bottom panel of Figure 1 shows the dust temperature for various
grain radii at a range of depths through the shell.  In Zone 1, we see
that only the largest grains with radii $a\ga$0.2 $\micron$ can match
the observed dust color temperatures.  The transition to higher
densities at the beginning of Zone 2 is accompanied by a sharp drop in
the grain temperature, as observed in mid-IR images (Smith et all.\
2003b).  This sharp drop to lower grain temperatures only occurred for
the highest-density models with n$_H$=10$^{6.7}$ to 10$^{7.0}$
cm$^{-3}$ in Zone 2.  The grain temperature in Zone 2 appears to be
independent of grain size.  This is perhaps because much of the
short-wavelength FUV radiation has been extinguished and the grains
are heated predominantly by radiation at near-UV, visual, and IR
wavelengths instead, where the absorption efficiency is less sensitive
to grain size (e.g., Draine 2003).  In this case, grain-gas
photoelectric heating will drop as well.  In fact, we see that the
grain temperature eventually drops to within 20 K of the equilibrium
blackbody temperature T$_{BB}$ (dashed line), independent of grain
radius.  The fact that the temperature in Zone 2 is nearly independent
of grain size is interesting, because it contradicts long-held
conventional wisdom.  The observed near-blackbody color temperature
has often been taken as an indication that the grains have radii of
1--2 $\micron$ (Smith et al.\ 2003b, 1998; Polomski et al.\ 1999;
Hackwell et al.\ 1986; Mitchel et al.\ 1983).  However, Figure 1
indicates that grains this large may not be necessary to explain the
grain temperature in Zone 2 after all (although some large grains are
still needed for the nearly gray extinction; Rodgers 1971; Hillier
2006; Whitelock et al.\ 1983).  In particular, a population that
includes {\it both} large and small grains is evidently allowed in
Zone 2.

The visual extinction through Zone 1 is only $\sim$1 mag, compared to
about 4.5--9 magnitudes for Zone 2 (Table 1).  Thus, the thin outer
H$_2$ shell dominates the visual extinction, the total infrared
luminosity, and also the scattered light seen in visual-wavelength
images and hard X-rays (Corcoran et al.\ 2004).  This is why model
shapes for the Homunculus derived from [Fe~{\sc ii}] or [Ni~{\sc ii}]
emission from Zone 1 (Davidson et al.\ 2001; Allen \& Hillier 1991)
are too small compared to images, whereas model shapes derived from
H$_2$ emission have the correct apparent size (Paper I).

The stellar continuum that illuminates the nebula is strongly
reprocessed by the inner region before it reaches the denser outer
zone.  Figure 3 shows the stellar continuum (the smoother dotted line)
that is incident upon the inner nebula.  The gray solid line with many
spectral emission features is the net emission that emerges from the
inner region and goes on to irradiate the molecular gas in Zone 2.
This continuum is a combination of the attenuated incident continuum
plus reprocessed emission from the inner nebula.  The UV continuum has
been heavily absorbed by dust within the inner region.  Thermal dust
emission dominates the infrared continuum, and the dust is warm enough
for the 10~$\micron$ silicate feature to come into emission.  Many
emission lines, mainly atomic and singly ionized species, are formed
in the inner region.  The major effect of this reprocessed continuum
will be upon dust in the outer nebula.  Much of the incident UV
continuum has been converted into thermal-IR emission already.  The UV
is the most effective in photoelectrically heating gas, so the outer
nebula is cooler as a result of the absence of this light, as noted
earlier.  The very strong infrared continuum heats the outer nebula
indirectly, by first heating the dust, which then shares its energy
with the gas as the result of gas-dust collisions.

The emergent spectrum from Zone 2 with $n_H$=10$^{6.7}$ cm$^{-3}$ is
shown with the solid black line in Figure 3.  As expected, the
thermal-IR continuum is shifted toward longer wavelengths because of
the cooler dust, and the UV/visual light is more severely attenuated
than for Zone 1.  Interestingly, when we compare the spectrum from
Zone 1 alone to the emergent spectrum from Zone 2 (which includes the
emergent spectrum of Zone 1 that passes through Zone 2), we see that
the 10~$\micron$ silicate feature has a different shape --- the
silicate profile from Zone 2 is somewhat muffled, indicating that the
emission is beginning to turn optically thick.  In fact, when the
density is increased to $n_H$=10$^7$ cm$^{-3}$ in Zone 2 (dashed gray
line) the silicate feature turns clearly to self-absorption.  This
trend in the silicate profile shape has two important implications.
First, when the emission begins to turn optically thick, the silicate
profile appears wider and flatter, and the peak moves toward longer
wavelengths.  This effect will modify the required contribution from
other grain species like corrundum (see Chesneau et al.\ 2005).
Second, the strong silicate feature observed in $\eta$ Car's mid-IR
spectrum may indicate that densities of $n_H$=10$^7$ cm$^{-3}$ are too
high (although the effects of clumping are an issue).  If densities
near our lower bound $n_H$=10$^{6.7}$ cm$^{-3}$ are preffered, then a
total mass of the Homunculus near 16 M$_{\odot}$ is favored over a
much larger value of $\sim$30 M$_{\odot}$ (see \S 5.1).  This is,
perhaps, a relief.  High spatial resolution spectroscopy of the
silicate profile in the Homunculus could provide a powerful and
independent constraint on the density and total mass of the nebula.

\section{CONDITIONS IN THE EQUATORIAL PLANE}

Near-infrared H$_2$ emission is seen at all latitudes in the
Homunculus (Paper I), even though the distance from the star changes
by an order of magnitude.  In the side walls of the polar lobes at
mid-latitudes, the oblique incident angle of the radiation
(i.e. ``seasons'') will mitigate the stronger radiation field at
closer radii to the central star, but this geometric advantage
disappears for the molecular gas at the pinched-waist of the nebula.
Thus, we also investigate the physical conditions in the equatorial
plane of the Homunculus, where H$_2$ is seen nearest to the star.  The
equatorial model was identical to the polar lobe/line-of-sight model,
except that the radii and thicknesses of Zones 1 and 2 were adjusted
to match observations (Paper I)\footnote{Note, however, that the
radius at the equator depends on azimuth, due to large-scale irregular
structures in the inner torus (Smith et al.\ 2002; Chesneau et al.\
2005; Paper I).}, and the density in Zone 1 was increased.  The
equatorial model is summarized in Table 2.

Figure 4 shows the results of a {\sc cloudy} calculation matched to
observations of the radial stratification in the equatorial plane of
the Homunculus (Paper I). The inner radius is set approximately equal
to the distance between the star and the Weigelt knots, which is
0$\farcs$2 to 0$\farcs$3 (Smith et al.\ 2004; Dorland et al.\ 2004),
or roughly 10$^{16}$ cm if we correct for the 41\arcdeg\ inclination
angle (Paper I; Davidson et al.\ 2001).  In order to match observed
electron densities of $>$10$^6$ cm$^{-3}$ in the Weigelt knots and
other bright equatorial ejecta (Smith 2002; Hamann et al.\ 1999, 1994;
Davidson et al.\ 1995), the simulations required a hydrogen density of
about n$_H$=2$\times$10$^6$ cm$^{-3}$ or more.\footnote{This is only
the average density.  In fact, some clumps like the cores of the
Weigelt knots may have much higher peak densities of 10$^8$ to 10$^9$
cm$^{-3}$ (Hamann et al.\ 1999).}  This was also the highest density
that allowed H to remain predominantly atomic and Fe to remain as
Fe$^+$ throughout Zone 1.  Even though the material is much closer to
the star and feels a much stronger radiation field than in the polar
model, $n_H\simeq$10$^7$ cm$^{-3}$ in Zone 2 was still sufficient to
allow a sharp transition to molecular hydrogen and neutral Fe.

Unlike the polar model, the inner edge of Zone 1 experienced a more
substantial H$^+$ fraction of $\sim$75\%, while $\sim$20\% of the iron
was ionized to Fe$^{++}$.  This ionization occurs despite the fact
that, aside from hard X-rays (Corcoran et al.\ 2001), we did not
include radiation from a hot companion star in our
calculation. Throughout most of Zone 1, the typical molecular hydrogen
fraction was of order 10$^{-13}$, while the neutral Fe fraction was
typically 10$^{-6}$.

The bottom panel of Figure 4 shows the dust temperature for grains in
the equatorial model.  In Zone 1, the predicted dust temperatures
greatly exceed observational constraints on the color temperature for
most grain sizes.  The hottest extended dust in the core of the
Homunculus (associated with the Weigelt knots and more distant ejecta)
is about 550 K (Smith et al.\ 2003b; Chesneau et al.\ 2005).  This is
marginally consistent with $a$=0.2 $\micron$ grains within
observational uncertainty, but the observed tempertures would be more
consistent with significantly larger grains.  The transition to lower
dust temperature at the boundary between Zones 1 and 2 is not as stark
as in the polar model, but like the polar model, the grain temperature
in Zone 2 drops and becomes insensitive to grain size (Fig.\ 4).  Here
it fully reaches temperatures as low as the equilibrium blackbody
temperature T$_{BB}$.

Figure 5 shows the incident (dotted) and transmitted energy
distributions for the equatorial model.  The transmitted radiation is
that which escapes from the outer boundary of Zone 2.  This emission
at the equator is of particular interest, as it illuminates material
in the equatorial skirt at larger radii.  Some of that equatorial gas
has pecular spectral properties, like the so-called ``strontium
filament'' (Hartman et al.\ 2004; Zethson et al.\ 2001; Bautista et
al.\ 2006, 2002), the very bright and possibly fluorescent [Ni~{\sc
ii}] $\lambda$7379 emission (Davidson et al.\ 2001), and He~{\sc i}
$\lambda$10830 emission (Smith 2002).  In Figure 5 we see that
radiation shortward of 2000 \AA\ is severely attenuated, while near-UV
radiation beyond this limit remains strong.

\section{DISCUSSION}

\subsection{Threshold Density for H$_2$ Survival}

Our most important observational constraint on the density in the
polar lobes is that molecular hydrogen is able to survive and Fe
becomes neutral, even though they are within only 0.1 pc of the most
luminous hot supergiant in our Galaxy.  Furthermore, both of these
conditions should be met in the polar model of Zone 2 with some room
to spare, since H$_2$ emission is seen to be fairly uniform at all
latitudes, including lower latitudes closer to the equator (and closer
to the star) where the radiation field is stronger.  As we have
discussed above, these constraints are only met for high densities of
n$_H$=10$^{6.7-7.0}$ cm$^{-3}$ in Zone 2.

From the thermal-IR spectral energy distribution and conservative
assumptions, Smith et al.\ (2003b) derived a lower limit to the total
mass of the Homunculus of about 12.5 M$_{\odot}$ (about 11 M$_{\odot}$
in the cool 140 K shell, and 1.5 M$_{\odot}$ in the warmer inner
shell).  If this mass of cool dust were spread evenly over the volume
of the H$_2$ shell inferred observationally, the density would be at
least n$_H$=10$^{6.5}$ cm$^{-3}$ in Zone 2 (Paper I).  However, our
{\sc cloudy} calculations show that even higher densities are likely,
as noted above.  These higher average densities distributed over the
same volume would imply a total gas mass for the Homunculus in the
range of roughly 15--35 M$_{\odot}$.  Thus, our study confirms the
large mass of the Homunculus deduced from IR observations of dust, but
uses an independent method sensitive to the gas density.  It also
confirms that previous IR estimates were indeed lower limits to the
total mass.  We also find that 35 M$_{\odot}$ is an upper limit to the
mass, since for densities much higher than 10$^7$ cm$^{-3}$, the
transition to molecular gas would occur earlier and the H$_2$ shell
would be thicker than observed.

Caveats: Although our mass estimate is sensitive to the gas density,
it is not completely independent of the gas:dust ratio, since dust
helps attenuate the UV radiation in the Lyman-Werner bands that
destroys H$_2$, and grains act as a catalyst by providing a surface
for H$_2$ formation.  However, like mass estimates from thermal-IR
dust emission, these are probably underestimates if they are wrong.

\subsection{Dust Temperature and Grain Properties}

Our {\sc cloudy} calculations were tuned to match the observed
structure in IR lines of H$_2$ and [Fe~{\sc ii}], but they naturally
reproduce the bimodal stratification of the observed dust color
temperature as well.  High-resolution thermal-IR images show a thick
inner shell at $\sim$200 K and a thin outer shell at $\sim$140 K
(Smith et al.\ 2003b), where the cooler dust color temperatures are
coincident with the outer H$_2$ shell.  This is clearly reproduced in
our model (Fig.\ 1).

Our earlier simulations with {\sc cloudy} used a uniform density of
10$^6$ cm$^{-3}$ throughout a thick shell (Ferland et al.\ 2005) and
were able to reproduce a transition from H to H$_2$ and Fe$^+$ to
Fe$^0$ at some point within the shell, {\it but in that model a sharp
drop in dust temperture did not occur at the same position as the
onset of H$_2$}.  Instead, in the constant-density thick-shell model,
the dust temperature began to drop immediately at the inner radius of
the shell and continued smoothly thereafter.  Therefore, the observed
sharp transition from a geometrically thick zone of $\sim$200 K to a
very thin layer of $\sim$140 K dust coincident with the onset of the
thin H$_2$ shell (Fig.\ 1) gives a robust and independent confirmation
that a sharp density contrast between the two zones is indeed
required.

Because of the strong density increase of a factor of 15-30 from Zone
1 to Zone 2, the model also successfully reproduced the relative
contributions to the total mass observed in each component.  Fits to
the IR spectral energy distribution (Smith et al.\ 2003b) show that
the 200 K dust in the inner shell only contains about 10\% of the
total mass in the nebula, while the rest of the mass resides in the
thin outer shell at 140 K.  Because the outer shell is so thin, it
must be much denser in order to have a higher total mass.  The
relative masses for the inner ($\sim$1 M$_{\odot}$) and outer (16-33
M$_{\odot}$) shells in Table 1 agree with these observational
constraints.  By stark contrast, a uniform-density thick shell would
have much more mass in the [Fe~{\sc ii}] zone.

Interestingly, the radial behavior of the dust temperature brings into
question long-held notions about the grain properties.  Several
authors have noted that the low dust color temperatures at thermal-IR
wavelengths are close to the equilibrium blackbody temperatures at
those radii, suggesting large grain sizes above $a\simeq$1~$\micron$
(Mitchell et al.\ 1983; Mitchell \& Robinson 1986; Hackwell et al.\
1986; Robinson et al.\ 1987; Apruzese 1975; Smith et al.\ 1998, 2003b;
Polomski et al.\ 1999).\footnote{Note, however, that Mitchell et al.'s
suggestion of very large was grains was also motivated in part by the
very broad 10~$\micron$ silicate emission feature, but Chesneau et
al.\ (2005), Mitchell \& Robinson (1978), and Hyland et al.\ (1979)
have suggested that this may be due to a partial contribution from
other grain emission features like corrundum.}  In contrast, we find
that the dust temperature in the outer layer is {\it independent of
grain size} (Figs.\ 1 and 4), and approaches the equilibrium blackbody
temperature even for small grains.  The material becomes optically
thick and severely attenuates the UV radiation, leaving the burden of
energy balance to heating by longer wavelength radiation.  At longer
wavelengths, the difference between absorption efficiency and the
thermal-IR emissivity is smaller, allowing small and large grains to
maintain similar temperatures.  The dust temperature never gets below
the equilibrium blackbody temperature in our calculations.  Therefore,
the very low dust temperature of 110 K that Morris et al.\ (1999)
deduced from fits to the far-IR spectral energy distribution is
probably too low, especially at the small radii in the equator
suggested by those authors.

Although the temperature is apparently independent of grain size in
Zone 2, very small grains are not permitted in Zone 1 because
corresponding high temperatures are not seen there.  The color
temperature in Zone 1 is about $\sim$200 K, which would be consistent
with grains of radius $a\ga$0.2~$\micron$.  Thus, our calculations
relax but do not eliminate the constraints on large grain size posed
by the low observed dust temperatures in the Homunculus.  Allowing the
presence of somewhat smaller grains may be relevant to the high degree
of polarization (Thackeray 1961; Visvanathan 1967; Warren-Smith et
al.\ 1979; Meaburn et al.\ 1987; 1993; Walsh \& Ageorges 2000) and the
unusual scattering properties of the Homunculus seen in
high-resolution polarization imaging (Schulte-Ladbeck et al.\ 1999).

\subsection{Bright [Fe~{\sc ii}] Emission from the Inner Shell and the Gas-phase Iron Abundance}

An accurate model for Zone 1 determines the radiation field that
penetrates it to illuminate the outer H$_2$ shell (Fig.\ 3), but the
inner shell is interesting in its own right because of the physical
conditions that give rise to the extraordinarily bright infrared
[Fe~{\sc ii}] emission seen there (Paper I).  These bright infrared
[Fe~{\sc ii}] emission lines are a common feature of other LBV nebulae
(Smith 2002b; Smith \& Hartigan 2006), which is in contrast to the
outer H$_2$ shell that seems to make $\eta$~Car unique among hot
luminous stars.

To investigate this bright [Fe~{\sc ii}] emission in detail, we ran
{\sc cloudy} using the more sophisticated large Fe atom.  This
predicts the strength of numerous Fe lines, including the distinct
[Fe~{\sc ii}] ($a^6D-a^4D$) and ($a^4F-a^4D$) transitions in the 1--2
$\micron$ region.  For example, with solar Fe abundance, our {\sc
cloudy} model for Zone 1 predicts a total luminosity for [Fe~{\sc ii}]
$\lambda$16435 of about 65 L$_{\odot}$.  This line has been observed
extensively in $\eta$~Car in both moderate and high resolution spectra
(Smith 2002a; Paper I), and exhibits an average brightness in the
polar lobes of about 2$\times$10$^{-12}$ ergs s$^{-1}$ cm$^{-2}$
arcsec$^{-2}$ (although the local value is highly spatially
dependent).  Integrating this over the roughly 140-150 arcsec$^2$
projected area of the Homunculus would indicate a total flux in this
line of 3$\times$10$^{-10}$ ergs s$^{-1}$ cm$^{-2}$ (excluding the
bright [Fe~{\sc ii}] contribution from the Little Homunculus; Smith
2005).  This corresponds to a total observed [Fe~{\sc ii}]
$\lambda$16435 luminosity of about 40--50 L$_{\odot}$.  The agreement
between our model's predicted luminosity of 65 L$_{\odot}$ and the
observed value of $\sim$45 L$_{\odot}$ in this same line is
remarkable, considering that the bipolar Homunculus has a somewhat
smaller projected area on the sky than the equivalent-radius spherical
shell in {\sc cloudy}.  This match indicates that the gas-phase Fe
abundance is indeed roughly solar in Zone 1.

A solar gas-phase Fe abundance here is actually quite astonishing.  In
dusty regions of the ISM, the abundance of Fe is normally depleted by
two orders of magnitude due to the formation of Fe-bearing grains
(e.g., Shields 1970).  We don't expect the overall dust-to-gas ratio
in Zone 1 to be unusually low, since the mass inferred for the gas
from the observed electron density agrees with the mass measured from
the 200 K dust component that occupies the same region (uncertainties
of a factor of 2 are likely, but not factors of 100).  Therefore, some
process has selectively destroyed Fe-bearing grains {\it without}
destroying the remaining grain population, or some mechanism has
specifically inhibited the formation of Fe-bearing grains.

Shock speeds $\ga$80 km s$^{-1}$ can destroy iron grains and release
the corresponding atomic Fe into the gas phase.  While we see no
spectroscopic evidence for such fast shocks in the polar lobes today,
shocks of this speed may have been active earlier in the life of the
Homunculus.  If such shocks destroyed the Fe-bearing grains early-on,
they may have been released into an environment where the density was
too low to reform the grains, locking Fe in the gas phase.
Alternatively, the ejecta in the Homunculus are C- and O-poor, and
N-rich.  The low oxygen abundance may have inhibited the formation of
iron oxides, which can be an important repository for storing Fe in
the solid phase of the ISM.  In the absence of C, O, and Fe, grains
might form preferentially with Al, Mg, or Ca instead.  Interestingly,
mid-IR spectroscopy of the Homunculus strongly suggests the presence
of Al-bearing grains like corrundum (Mitchell \& Robinson 1978; Hyland
et al.\ 1979; Chesneau et al.\ 2005).  We also know that Fe is present
in the gas phase in the outer H$_2$ shell, since Fe$^0$ is seen in
absorption there in UV spectra (Gull et al.\ 2005; Nielsen et al.\
2005).  The existence of a region of the ISM with a high dust content
and where Fe is predominantly in the gas phase has important
ramifications beyond the study of $\eta$ Carinae, and will be
discussed elsewhere.

\subsection{Origin of the Double-Shell Structure}

We have shown that the observed double-shell structure of the
Homunculus requires a strong density jump in the walls of the nebula.
The observed transition from atomic to molecular gas (Paper I) and the
corresponding drop in dust temperature (Smith et al.\ 2003b) result
when FUV radiation propagating through the low-density inner zone
suddenly encounters the high densities in the thin outer shell.  The
absorption of this UV radiation in the Lyman-Werner bands by gas and
dust in the dense outer shell regulates the destruction rate of H$_2$,
allowing the H$_2$ to survive.  The required density contrast between
these two zones is a factor of roughly 15--30.

While our {\sc cloudy} calculations explain how the observed
properties of the double-shell structure result from a strong density
stratification, they do not explain the origin of the density
structure itself.  Since the Homunculus appears to be expanding
ballistically (Smith \& Gehrz 1998) and because the current stellar
wind is too weak to shape the massive nebula, it is likely that this
density structure was determined early.  The observed stratification
is reminiscent of the layered structure seen in cooling zones behind
shock fronts; although this is unlikely to be occuring at the present
time (Paper I), the observed density structure may be frozen-in to the
ejecta from shocks that occurred during the event itself as the dense
material cooled.  As noted in the previous section, shocks present
during an earlier epoch would also be interesting from the point of
view of the gas-phase Fe abundance.  The rapid cooling of this dense
layer may have triggered severe thermal instabilities, which in turn,
could explain the currently-observed clumping and fragmentation seen
in images of the Homunculus (Morse et al.\ 1998).  Alternatively, the
low-density inner zone may result from a photoevaporative flow off the
inside face of the much denser outer H$_2$ shell, as is commonly seen
in photodissociation regions at the surfaces of giant molecular
clouds.  Finally, an obvious possibility is that the two shells simply
result from separate ejection events during the $\sim$20 yr timespan
of the Great Eruption.  We know that this is at least plausible from
the existence of the outer ejecta, the Homunculus, and the Little
Homunculus, indicating that $\eta$ Car has suffered multiple mass
ejections (Walborn 1976; Walborn et al.\ 1978; Ishibashi et al.\ 2003;
Smith 2005).

Our models that reproduce the fundamental observed structure have
constrained the most basic physical properties of the Homunculus:
namely, its density and temperature.  These will allow further
refinements to models trying to understand emission line spectra,
unusual grain properties, molecular chemistry, abundances, and
radiative transfer models of the appearance of $\eta$ Car's nebula in
images.  The physical explanation for this distinct double-shell
structure may have wider applications as well, especially to objects
like planetary nebulae.  For example, the bipolar nebula M2-9 has a
similar and pronounced double-shell structure, with a thin outer H$_2$
shell and an inner [Fe~{\sc ii}] shell (Smith et al.\ 2005; Hora \&
Latter 1994).  The UV radiation field of M2-9 is known to be highly
asymmteric and time variable because of a central binary system (e.g.,
Doyle et al.\ 2000), so it is likely that a density jump controls the
shape of the H$_2$ lobes, as in $\eta$ Car.  Our {\sc cloudy} models
for $\eta$ Car would then suggest that the transition from atomic to
molecular gas in such PNe would be accompanied by a similar sharp drop
in observed dust color temperature at mid-IR wavelengths, possibly
insensitive to grain size as well.

\subsection{Future Evolution of the H$_2$ Shell}

Since $\eta$ Car is the only LBV nebula known to have a dense H$_2$
shell, we must ask if this is a truly unique property of $\eta$ Car
itself, or if it is simply a consequence of $\eta$ Car being observed
at a special time so soon after its mass ejection; the Homunculus is
also the {\it youngest} LBV nebula.  We expect that in the decades
immediately following the outburst, the ejecta around $\eta$ Car were
extremely dense and self-shielding, permitting the formation of not
only dust, but also H$_2$ and even more complex polyatomic molecules
like ammonia (Smith et al.\ 2006).  The physical conditions in this
early high-density epoch are of great interest.  This material has
since been expanding and thinning, as the observed extinction drops
and the star brightens (Whitelock et al.\ 1983; Davidson et al.\ 1999;
Smith et al.\ 2000).

Due to the high mass of the Homunculus and its relatively empty
surroundings, the shell will continue to expand quasi-ballistically.
Thus, the hydrogen density will drop as $r^{-2}$, and because of
ballistic motion, roughly as $t^{-2}$ as well. In another couple
hundred years, the density in the nebula will drop below the threshold
at which H$_2$ can survive (this may happen even sooner, since $\eta$
Car resides in a giant H~{\sc ii} region with a strong ambient UV
field). Eventually, the Homunculus will become an atomic shell
resembling the massive shells seen around other LBVs and LBV
candidates, like the more evolved nebulae around the Pistol star
(Figer et al.\ 1999), AG Car (Voors et al.\ 2000), and P Cygni (Smith
\& Hartigan 2006).  At these later stages, the Homunculus may even
overtake and mix with more distant outer ejecta (Walborn 1976),
appearing as a single ring-like LBV nebula, thereby erasing the
details of its eruptive history.

\section{CONCLUSIONS}

We have modeled the basic double-shell structure in the Homunculus, as
traced most vividly by near-IR emisison from [Fe~{\sc ii}] and H$_2$
(Paper I). Our aim was to constrain the density which permits the
survival of H$-2$, and thus, to provide an independent constraint on
the mass of the Homunculus and the physical conditions in the
nebula. The main observational constraints on our {\sc cloudy} models
were the geometric thickness of each zone, the electron density in the
inner [Fe ~{\sc ii}] zone, and the survival of both H$_2$ and Fe$^0$
in the thin outer zone.  Our primary conclusions are as follows:

1. The hydrogen density in the outer H$_2$ zone is
   $n_H\simeq$(0.5--1)$\times$10$^7$ cm$^{-3}$.

2. Using this density and the observed geometry of the nebula, the
   total mass is roughly 15-35 M$_{\odot}$, and we note reasons why
   the lower end of this range is favored.

3. The relative mass of gas in the two zones closely matches that
   derived for warm and cool dust components in the same two zones
   (Smith et al.\ 2003b).  This suggests that previous assumptions of
   a normal gas:dust mass ratio of 100 are not wildly in error.

4. We predict a sharp drop in the dust temperature between the inner
   and outer zones of the double shell, as observed (Smith et al.\
   2003b).  The temperature in the outer zone approache the
   equilibrium blackbody temperature independent of grain size, due to
   shielding of UV radiation.  This relaxes constraints on the
   unusually large grains inferred from observed dust temperatures.

5. Even though dust formation was efficient in the ejecta of $\eta$
   Car, we find that the gas phase Fe abundance is still roughly
   solar.  One normally expects Fe to depleted by two orders of
   magnitude in dusty environments.  This has occurred despite the
   severe depletion of C and O in the ejecta.

\acknowledgments \scriptsize

N.S.\ was supported by NASA through grant HF-01166.01A from the Space
Telescope Science Institute, which is operated by the Association of
Universities for Research in Astronomy, Inc., under NASA contract
NAS5-26555.  Research into the physical processes in the ISM by
G.J.F.\ is supported by NSF (AST0307720) and NASA (NAG5-12020).


\begin{deluxetable}{lccccccc}
\tabletypesize{\scriptsize}
\tighten\tablenum{1}\tablewidth{0pt}
\tablecaption{Calculation of the Double-shell structure at the Pole}
\tablehead{
  \colhead{Parameter} &\colhead{Units} &\colhead{Zone 1}
  &\colhead{Zone 2}  &\colhead{Zone 2} &\colhead{Zone 2} &\colhead{Zone 2} &\colhead{Zone 2}
}
\startdata

\cutinhead{Input parameters}

log R$_{in}$		&cm		&17.255	&17.398	&17.398	&17.398	&17.398	&17.398	\\
log R$_{out}$		&cm		&17.398	&17.415	&17.415	&17.415	&17.415	&17.415	\\
log $\Delta$R		&cm		&16.845	&16.0	&16.0	&16.0	&16.0	&16.0	\\
log n$_H$   		&cm$^{-3}$	&5.5	&7.0	&6.7	&6.5	&6.3	&6.0	\\
log N$_{H}$ 		&cm$^{-2}$	&22.35	&23.0	&22.7	&22.5	&22.3	&22.0	\\
M$_{total}$\tablenotemark{a} &M$_{\odot}$ &1	&33	&16	&11	&7.5	&3.7	\\
filling factor		&\nodata	&0.5	&0.9	&0.9	&0.9	&0.9	&0.9	\\

\cutinhead{Calculated parameters}

log n$_e$		&cm$^{-3}$	&3.7	&4.5	&4.35	&4.26	&4.2	&4.0	\\
T$_{\rm e}$		&K		&170	&118	&142	&169	&190	&191	\\
Fe$^+$/Fe		&\nodata	&1.0	&2(-4)	&0.97	&1.0	&1.0	&1.0	\\
n(H$^+$)/n$_H$		&\nodata	&0.016	&2(-12)	&2(-3)	&3(-3)	&3(-3)	&1(-4)	\\
max 2n(H$_2$)/n$_H$ 	&\nodata	&5(-10)	&1.0	&1.0	&0.99	&0.42	&2(-4)	\\
D 2n(H$_2$)/n$_H$=0.5	&...		&...	&0.15	&0.36	&0.61	&...	&...	\\
D 2n(H$_2$)/n$_H$=0.99	&...		&...	&0.41	&0.88	&...	&...	&...	\\
$A_V$			&mag		&1.1	&8.9	&4.5	&2.8	&1.8	&0.9	\\

\enddata

\tablecomments{The first seven rows are prescribed inputs.  The
results of the calculation in the remaining rows are example values at
roughly the midpoints in each zone.  The value of n$_H$ corresponds to
the total hydrogen density n(H$_2$+H$^0$+H$^+$).  In the last two
rows, the value of D gives the fractional depth into the shell at
which the H$_2$ fraction reaches either 50 or 99 \%.}

\tablenotetext{a}{This is the total mass assuming that the volume of
  the bipolar Homunculus (Paper I) is filled with the shell's density.
  This mass will be differrent from that for an equivalent spherical
  shell.}
\end{deluxetable}

\begin{deluxetable}{lccc}
\tabletypesize{\scriptsize}
\tighten\tablenum{2}\tablewidth{0pt}
\tablecaption{Double-shell structure at the Equator}
\tablehead{
  \colhead{Parameter} &\colhead{Units} &\colhead{Zone 1}  &\colhead{Zone 2}
}
\startdata

\cutinhead{Input parameters}

log R$_{in}$		&cm		&16.0	&16.498		\\
log R$_{out}$		&cm		&16.498	&16.618		\\
log $\Delta$R		&cm		&16.332	&16.0		\\
log n$_H$   		&cm$^{-3}$	&6.3	&7.0		\\
log N$_{H}$ 		&cm$^{-2}$	&22.63	&23.0		\\
filling factor		&\nodata	&0.5	&0.9		\\

\cutinhead{Calculated parameters}

log n$_e$		&cm$^{-3}$	&6.2	&5.7		\\
T$_{\rm e}$		&K		&610	&170		\\
Fe$^+$/Fe		&\nodata	&0.82--1 &4(-4)	\\
n(H$^+$)/n$_H$		&\nodata	&0.74	&2(-12)	\\
max 2$\times$n(H$_2$)/n$_H$ &\nodata	&2(-13)	&1.0	\\

\enddata

\tablecomments{The first six rows are prescribed inputs.  The results
of the calculation in the remaining rows are example values at roughly
the midpoints in each zone.  The value of n$_H$ corresponds to the
total hydrogen density n(H$_2$+H$^0$+H$^+$).}
\end{deluxetable}

\begin{figure}
\epsscale{0.7}
\plotone{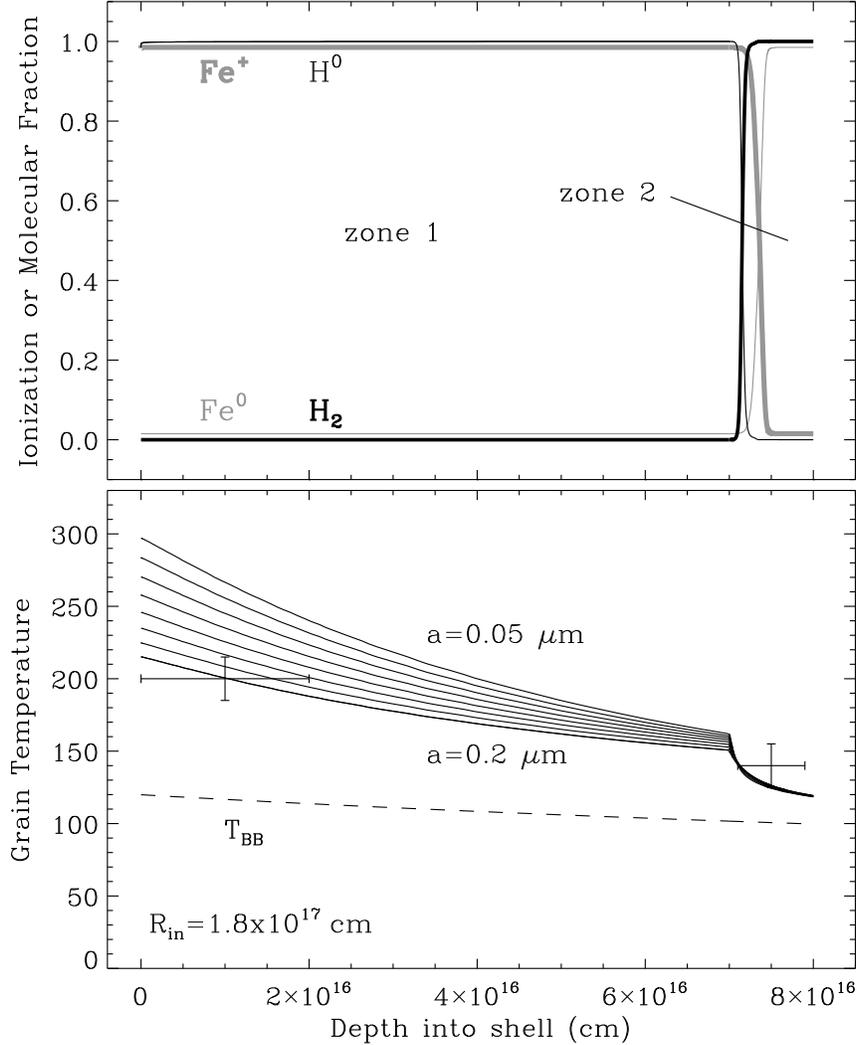} 
\figcaption{Variation of observable physical properties throughout the
thick shell, where the left edge marks the inner radius of the inner
shell (zone 1), and the right edge marks the outer surface of the
Homunculus.  The top panel shows the variation in the relative
fraction of atomic/molecular hydrogen (thin/thick black lines), as
well as the neutral and singly-ionized Fe fractions (thin/thick grey
lines).  Here we show the model with $n_H$=10$^7$ cm$^{-3}$ in Zone 2.
The bottom panel shows the grain temperature at the same locations for
a range of different grain radii $a$, where the dashed curve is the
equilbrium blackbody temperature.  The error bars in the bottom panel
show observed grain color temperatures for the inner and outer shells
(from Smith et al.\ 2003b).  The physical parameters in the model are
summarized in Table 1.}
\end{figure}

\begin{figure}
\epsscale{0.5}
\plotone{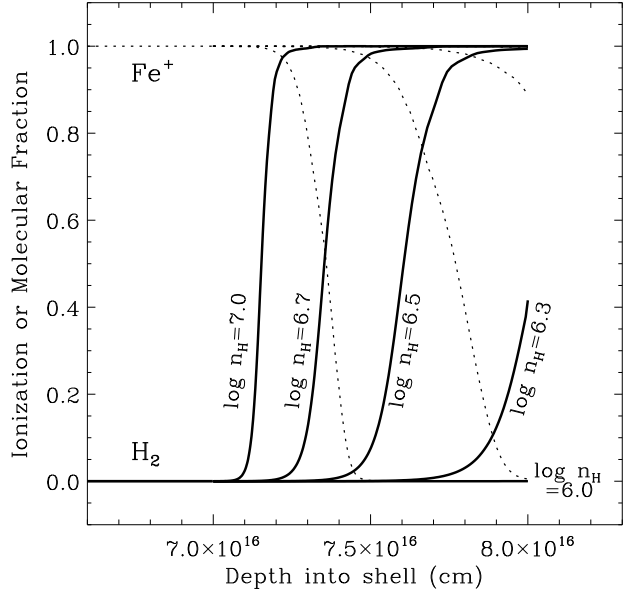} 
\figcaption{A detail of Figure 1, concentrating on Zone 2, showing the
  transition to molecular hydrogen for several different values of the
  density in Zone 2 (see Table 1).  The molecular gas fraction is
  shown by solid curves, while the ionization fraction of Fe$^+$ is
  shown by the dotted curves.}
\end{figure}

\begin{figure}
\epsscale{0.5} 
\plotone{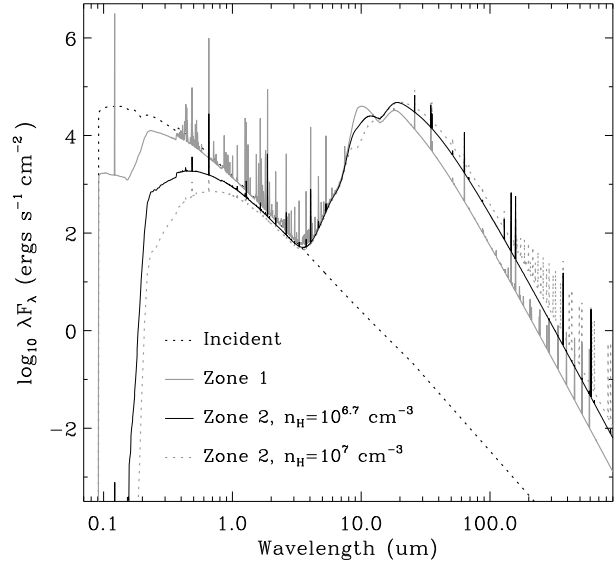} 
\figcaption{The incident continuum (dashed) and emergent spectra from
  Zone 1 (solid gray) and Zone 2 for the model in the polar lobes.
  For Zone 2 we show the emergent spectrum for both plausible hydrogen
  densities of 10$^{6.7}$ cm$^{-3}$ (solid black) and for 10$^7$
  cm$^{-3}$ (dashed gray line).}
\end{figure}

\begin{figure}
\epsscale{0.5}
\plotone{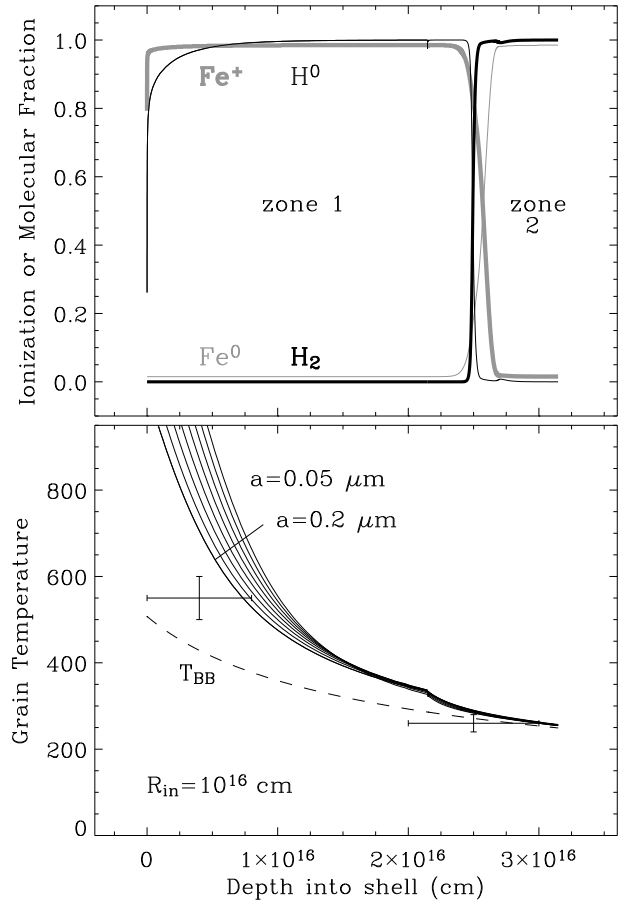} 
\figcaption{Same as Figure 1, but showing the model for the
  ionization/molecular fraction and dust temperature in the equatorial
  plane of the Homunculus (see Table 2).}
\end{figure}

\begin{figure}
\epsscale{0.5}
\plotone{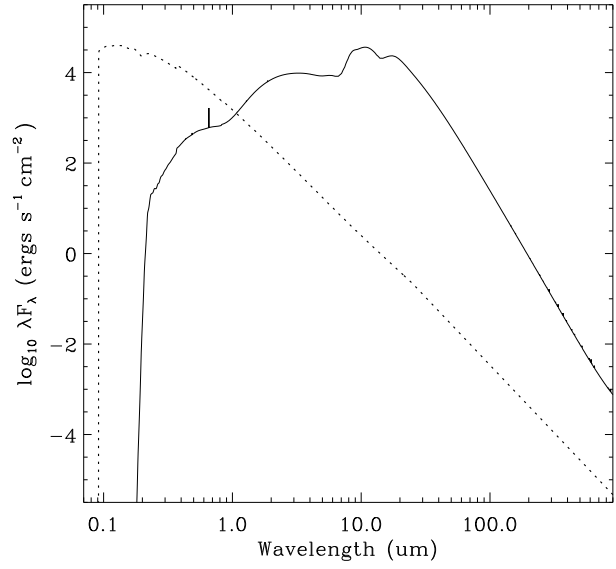} 
\figcaption{The incident (dashed) and transmitted (solid) continuum
  radiation for the model in the equatorial plane.  This transmitted
  continuum radiation escapes to illuminate gas and dust at larger
  radii in the equatorial skirt, like the so-called strontium
  filament.}
\end{figure}

\end{document}